\documentstyle[12pt,emulateapj,epsfig]{article} 
\newcommand{\kms}{km s$^{-1}$}       
\newcommand{\cmtwo}{cm{$^{-2}$}} 
\newcommand{\cmthree}{cm{$^{-3}$}} 
\newcommand{\um}{$\mu$m}                                 
\newcommand{\lsun}{$L_{\odot}$}               

\newcommand{\msunyr}{$M_{\odot}$ yr$^{-1}$}

\newcommand{\lsim}{\;\lower.6ex\hbox{$\sim$}\kern-7.75pt\raise.65ex\hbox{$<$}\;}
\newcommand{\gsim}{\;\lower.6ex\hbox{$\sim$}\kern-7.75pt\raise.65ex\hbox{$>$}\;}
\newcommand{\amin}{$^{\prime}$}                   
\newcommand{\asec}{$^{\prime \prime}$}

\newcommand{\lflux}{W cm$^{-2}$}

\newcommand{\oia}{[O{\sc i}]63\um}
\newcommand{\oib}{[O{\sc i}]145\um}
\newcommand{\cii}{[C{\sc ii}]158\um}
\newcommand{\sii}{[Si{\sc ii}]34.8\um}
\newcommand{\neii}{[Ne{\sc ii}]12.8\um}
\newcommand{\hii}{H$_2$}
\newcommand{\ho}{H$_2$O}
\newcommand{\bo}{${\bf B}_0$}
\newcommand{\bb}{{\bf B}}
\newcommand{\mbo}{$B_0$}
\newcommand{\vs}{${\bf v}_s$}
\newcommand{\vsj}{${\bf v}_{s_J}$}
\newcommand{\vsc}{${\bf v}_{s_C}$}
\newcommand{\mvs}{$v_s$}
\newcommand{\mvsj}{$v_{s_J}$}
\newcommand{\mvsc}{$v_{s_C}$}
\newcommand{\mug}{$\mu$G}
\newcommand{\ha}{H$\alpha$}
\newcommand{\suii}{[S{\sc ii}]6717+31\AA}
\begin{document}
%
%
%
\lefthead{Molinari et al.}
\righthead{ISO Spectroscopy of the HH\,7-11 Flow and its red-shifted 
counterpart}

\title{ISO Spectroscopy of the HH\,7-11 Flow and its red-shifted 
counterpart\footnotemark[0]}

\author{Sergio Molinari and Alberto Noriega-Crespo}
\affil{Infrared Processing and Analysis Center \& SIRTF Science Center, 
California Institute of Technology, MS 100-22, Pasadena, CA 91125, USA}
\author{Cecilia Ceccarelli}
\affil{Laboratoire d'Astrophysique, Observatoire de Grenoble - BP 53, 
F-38041 Grenoble cedex 09, France}
\author{Brunella Nisini, Teresa Giannini and Dario Lorenzetti}
\affil{Osservatorio Astronomico di Roma, via Frascati 33, I-00044 
Monte Porzio, Italy}
\author{Emmanuel Caux}
\affil{CESR CNRS-UPS, BP 4346, F-31028 Toulouse Cedex 04, France}
\author{Ren\'e Liseau}
\affil{Stockholm Observatory, S-133 36 - Saltsj\"osbaden, Sweden}
\author{Paolo Saraceno}
\affil{CNR-Istituto di Fisica dello Spazio Interplanetario, Area di 
Ricerca Tor Vergata, via Fosso del Cavaliere I-00133 Roma, Italy}
\author{Glenn J. White}
\affil{Department of Physics, Queen Mary and Westfield 
College, University of London, Mile End Road, London E1 4NS, UK}
\affil{Stockholm Observatory, S-133 36 - Saltsj\"osbaden, Sweden}

\footnotetext[0]{Based on observations with ISO, an ESA project
with instruments funded by ESA Member States (especially the PI
countries: France, Germany, the Netherlands and the United Kingdom) with
the participation of ISAS and NASA.}

\begin{abstract}

We have used the two spectrometers on the Infrared Space Observatory to
observe the HH\,7-11 flow, its red-shifted counterpart, and the
candidate exciting source SVS\,13, in the star formation region
NGC\,1333. We detect atomic (\oia, \oib, \sii, \cii) and molecular
(\hii, CO, H$_2$O) lines at various positions along the bipolar flow.

Most of the observed lines can be explained in terms of shock-excited
emission. In particular, our analysis shows that dissociative (J-type)
and non-dissociative (C-type) shocks are simultaneously present
everywhere along both lobes of the flow. We confirm the low-excitation
nature of the Herbig-Haro nebulosities, with shock velocities 
\mvs$\lsim 40-50$ \kms. Toward both lobes of the outflow we find 
pre-shock densities of $n_0 \sim 10^4$ \cmthree\ for both the J and C
components, implying \mbo$\sim$100\mug\ for \mbo$\propto n_0^{0.5}$. In
the central region of the flow, close to the exciting source, the
pre-shock density deduced for the  C-shock component is
$n_0\sim10^5$\cmthree, suggesting a magnetic  field $\sim$3 times
stronger. We propose that the deficiency of gas-phase water in the post
C-shock regions is due to freezing onto  warm grains processed through
the J-shock front and traveling  along the magnetic field lines. The
total observed cooling  from the dissociative shock components is
consistent with the power lost by a slow molecular outflow 
accelerated by a fast neutral H{\sc i} wind.

Finally, the skin of the cloud seen in projection toward the flow
appears to be weakly photo-ionised by BD +30$^{\circ}$\,549, the
dominant illuminating source of the NGC\,1333 reflection nebula.

\end{abstract}
\keywords{Stars: formation - (ISM:) Herbig-Haro objects - 
ISM: individual objects: HH\,7-11 - ISM: molecules - Infrared: ISM: 
lines and bands}
%

\section{Introduction}
\label{intro}

Atomic and molecular outflows trace the mass loss from protostellar 
objects, which is a fundamental characteristic of the formation and
evolution  of low mass stars. These outflows are often traced by the
optical Herbig-Haro (HH) objects, shock-excited nebulosities which mark
the interface between outflowing and circumstellar material. One of
these systems, which since its discovery  (Herbig~\cite{H74} ; Strom et
al.~\cite{Setal74}) has been subjected to a detailed multi-wavelength
analysis, is HH\,7-11. The system is relatively bright and lies
in the very active star forming NGC\,1333 region (Aspin et al.
~\cite{ASR94};  Bally et al.~\cite{Betal96}). A distance to the outflow
of 350pc (Herbig \& Jones~\cite{HJ83}) is widely adopted in the
literature, although Cernis (\cite{C93}) proposes 200pc. Early optical
spectroscopic studies show  that the HH\,7-11 outflow has a complex
velocity field and low excitation  (Solf \& B\"ohm~\cite{SB87}; B\"ohm
\& Solf ~\cite{BS90}), a conclusion  further supported by near infrared
studies (Hartigan et al.~\cite{HCR89};  Carr~\cite{Carr93}). Early on, a
well defined CO bipolar outflow was detected associated with this system
(Snell \& Edwards~\cite{SE81}; Bachiller \& Cernicharo~\cite{BC90}),
which also is observed in some other molecules like HCO$^+$ and H$_2$O
(Mehringer~\cite{M96};  Cernicharo et al.~\cite{CBG96}). 

The HH\,7-11 outflow has an unusual morphology; the blue outflow lobe is
made up of an arc-shaped chain of knots, while the red one is invisible
at optical wavelengths. The red lobe is, however, detected in the
(1-0)S(1) \hii\ line at 2.12\um\ having a very ragged appearance (see
Fig~\ref{map}). Both lobes have a total extension of $\sim $ 2\amin. It
has been thought that the driving source of the outflow is the infrared 
star SVS\,13 (Strom et al.~\cite{Setal76}), a conclusion partially
supported by  the proper motion measurements of the knots (Herbig \&
Jones\cite{HJ83}) and the source observed outbursts
(Goodrich~\cite{G86}; Eisl\"offel et al. ~\cite{Eetal91};  Liseau et
al.~\cite{LLM92}). The source has a luminosity of $\sim 85$ \lsun\
(Molinari et al.~\cite{MLL93}). Recent high angular resolution ($\sim$
0.\asec 3) VLA continuum observations at 3.6 cm suggest another nearby
embedded source (VLA\,3)  as a likely candidate, based on its better
alignment with the HH string. Interferometric observations (Bachiller et
al. ~\cite{Betal98}) at 1.3 and 3.5 mm with better than $0.\arcsec 2$
resolution,  however, have not confirmed this. The interferometric
observations have led to the discovery of a second jet emanating  from a
more deeply embedded source 14.\asec 5 away from SVS\,13, named SVS\,13B
(Grossman et al.~\cite{Getal87}).

One of the reasons why the HH\,7-11 flow has been so intensively studied
(see e.g. Reipurth~\cite{R94}), is that it was the first system showing
clear  signatures of a high velocity outflow in both neutral and
molecular gas tracers (Lizano et al.~\cite{Letal88}; Rodriguez et
al.~\cite{Retal90}; Giovanardi et al.~\cite{Getal92}).  This was a major
step forward in the interpretation of molecular outflows as being 
driven by faster but more tenuous (than the outflow entrained gas)
atomic stellar winds (Masson \& Chernin~\cite{MC93}; Raga et
al.~\cite{Retal93}). This has led to a more careful analysis of the 
energetics and shock conditions associated with the ionic/atomic and
molecular gas outflows (Raga~\cite{Ra91}). HH\,7-11 is also one of the
few examples where it is possible to disentangle the contributions of
shock excited and fluorescent emission  from its near infrared H$_2$
spectra (Gredel~\cite{G96}; Fernandes \& Brand~\cite{FB95};
Everett~\cite{E97}).

In the present study, we take advantage of the capabilities of the
Infrared Satellite Observatory (ISO, Kessler et al.~\cite{Ketal96})
spectrometers to study the mid- and far infrared emission line spectra
from the HH\,7-11 red and blue lobes and around the driving source
SVS\,13. The observations are described in Sect.~\ref{obs}; the results
are presented and discussed in Sect.~\ref{res} and following, and the
main conclusions are summarised in Sect.~\ref{summary}.

\section{Observations and Data Reduction}
\label{obs}

\begin{figure*}
\vspace{10.1cm}
\includegraphics{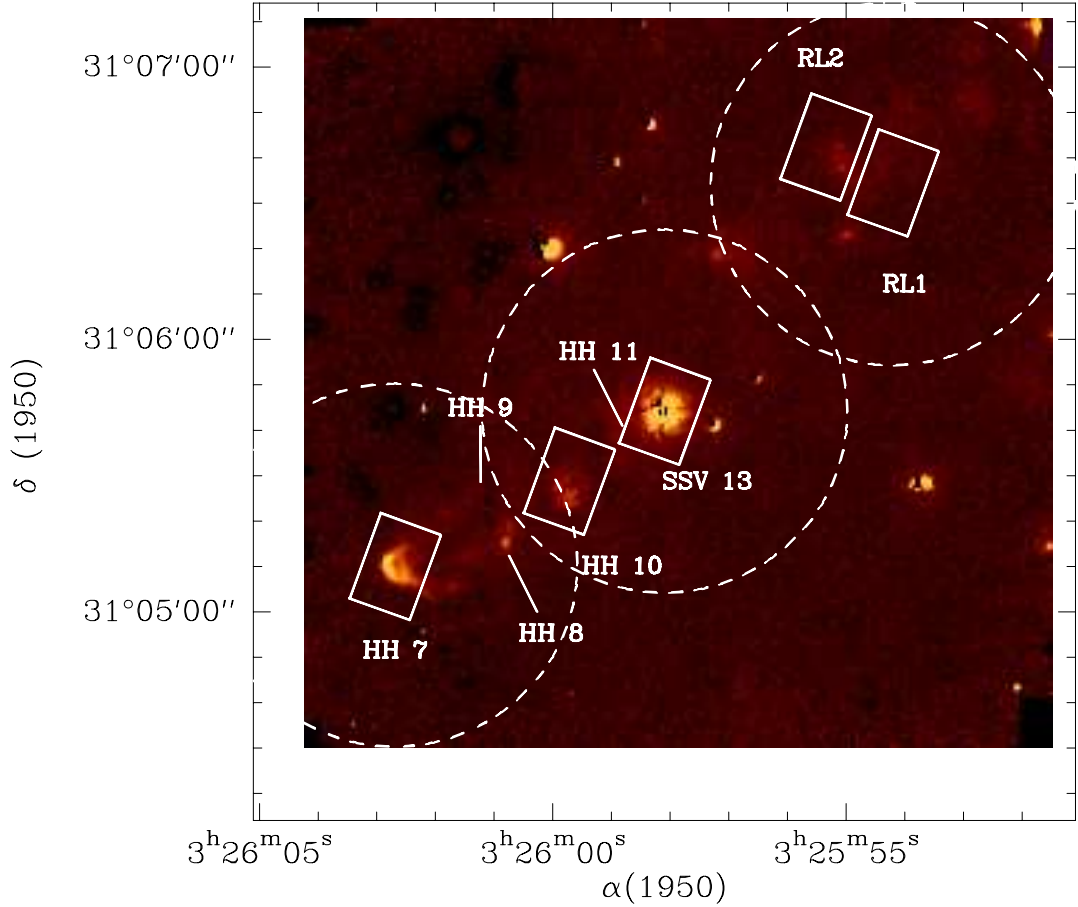}
\vspace{2cm}
\caption{\label{map} ISO fields of view in the
various observed positions, superimposed onto a continuum-subtracted
H$_2$ (1-0)S(1) line map. The dashed circles and solid rectangular
apertures represent the LWS and SWS  fields of view  centered on the
various observed positions. SVS\,13 is at the center of the middle LWS
(dashed circle) field  of view.}
\end{figure*}

\begin{table*}
\begin{center}
\caption{~~~~~~~~~~~~~~~~~~~~~~~~~~~~~~~~~~~~~~~~~~~~~~~~~~~~~~~~~~~Observations\label{obstab}}
\vspace{0.25cm}
\begin{tabular}{lcclcc}\tableline\tableline 
{Object} & {$\alpha$(1950)} & {$\delta$(1950)} & 
{AOT} & {Rev} & {Obs. Time}\\ \tableline 
SVS\,13 & 03 25 58.1 & $+$31 05 44.1 & LWS01 & 654 & 4053 \\
        &            &               & SWS02 & 814 & 1730 \\
HH\,7   & 03 26 02.7 & $+$31 05 10.2 & LWS01 & 654 & 3165 \\
        &            &               & LWS04 & 831 & 2428 \\
        &            &               & SWS02 & 652 & 1464 \\
HH\,10  & 03 25 59.8 & $+$31 05 28.8 & SWS02 & 847 & 1730 \\
Red Lobe 1 (RL\,1) & 03 25 54.3 & $+$31 06 34.1 & LWS01\tablenotemark{a} & 654 & 3165 \\
        &            &               & LWS04\tablenotemark{a} & 831 & 2428 \\
        &            &               & SWS02 & 652 & 1464 \\
Red Lobe 2 (RL\,2) & 03 25 55.4 & $+$31 06 42.1 & SWS02 & 847 & 1730 \\ 
\tableline
\end{tabular} \\
\vspace{-1cm}
\tablenotetext{}{~~~~~~~~~~~~~~~~~~~~~~~$^a$ The field of view also includes the position ``Red Lobe 2''}
\end{center}
\end{table*}

We used the two spectrometers on the ISO satellite to observe  several
locations along the HH\,7-11 flow, its optically invisible counterflow
and the candidate exciting source SVS\,13. The Long Wavelength
Spectrometer (LWS, Clegg et al.~\cite{Clegg96}) was used in its LWS01
grating mode to acquire full low resolution (R $\sim$ 200) 43-197\um\
scans with data collected every 1/4 of a resolution element (equivalent
to $\sim$0.07\um\ for $\lambda \lsim$90\um, and to $\sim$0.15\um\ for
$\lambda \gsim$90\um); a total of 19 scans were collected, corresponding
to 38s integration time per spectral element. The LWS was also used in
LWS04 Fabry-Perot (FP) mode to collect high (R $\sim$ 8000) resolution
scans of the \oia\ line; 52  scans sampled at 1/4 of the resolution
element (i.e., $\sim$ 0.0017 \um,  or 8 \kms) were collected, equivalent
to an integration time of $\sim$ 100s per spectral element. The Short
Wavelength Spectrometer  (SWS, de Graauw et al.~\cite{DGetal96}) was
used in its SWS02 grating mode to observe  line scans at medium
resolution (R $\sim$ 2000) for selected wavelength regions  covering the
\sii\ and \neii\ fine structure lines, and the pure rotational
transitions of molecular hydrogen from (0-0)S(1) to (0-0)S(7). All
relevant information is summarised in Table~\ref{obstab}, including the
Astronomical Observation Templates  used, the revolution number and the
total observing time of the observations.

LWS data processed through Off-Line Processing (OLP), version 7, have
been reduced using the LWS Interactive Analysis\footnote{LIA is
available at http://www.ipac.caltech.edu/iso/lws/lia/lia.html} (LIA)
Version 7.2. The dark current and gain for each detector were
re-estimated, and the data were recalibrated in wavelength, bandpass and
flux. The absolute flux  calibration quoted for LWS in grating mode is
10-15\% and it is valid  for point-like sources since the primary
calibrator, Uranus, is  point-like to the LWS beam; however, our sources
are not rigorously point-like and we adopt a more conservative number of
20\%. Additional processing for the FP data (with the LIA routine 
FP\_PROC) included a gain correction to compensate for the incorrect 
positioning of the grating during the FP observations. The integrated
\oia\ line fluxes obtained with the FP are about a factor of 2 lower
than the values measured using the grating. This discrepancy is larger
than the 30\% figure generally quoted for the absolute flux calibration
of the FP (Swinyard et al~\cite{Swin98}). We cannot offer any
explanation for  this difference, and in this paper we will not use the
line fluxes measured with the FP. The accuracy of the FP wavelength
calibration is believed to be better than 1/2 resolution element, or
$\sim$ 15 \kms.

SWS data were processed using OSIA, the SWS Interactive
Analysis\footnote{OSIA is available at
http://www.mpe.mpg.de/www\_ir/ISO/observer/osia/osia.html }. Dark
currents and photometric checks were revised; in many cases the former
were corrupted on a few detectors and were re-estimated and subtracted.
The March 1998 bandpass calibration files have been used to produce the
final spectra. The absolute flux calibration for SWS data should be accurate to within  20\%.

The final steps of data analysis were done using the ISO  Spectral
Analysis Package\footnote{ISAP is available at
http://www.ipac.caltech.edu/iso/isap/isap.html} (ISAP)  Version 1.5 for
both LWS and SWS. Grating scans (LWS) and detectors spectra (SWS) were
averaged using a median clipping algorithm optimised to flag and discard
outliers mainly due to transients; line fluxes were estimated by means
of gaussian  fitting (multiple gaussians in case of blended lines). The
LWS observations toward HH\,7 and the red lobe (RL\,1+2) were heavily
fringed due to the vicinity of the relatively strong continuum source
SVS\,13; standard techniques available under ISAP were used to remove
these instrumental effects.

The locations of the observed positions and instrument apertures are
shown in Fig.~\ref{map} superimposed on an H$_2$ (1-0)S(1) 2.12\um\
continuum-subtracted map  obtained with the near infrared camera at the
60\asec\ Mt. Palomar telescope (Murphy et al.~\cite{Metal95}). The
apertures of the two ISO spectrometers are quite different. The SWS
focal plane aperture is 14\asec x20\asec\ for all of the detected lines
except for \sii\ (20\asec x33\asec) so that, at each pointing, the 
contamination from nearby knots should be  negligible. The LWS aperture
is rather large ($\O\sim$80\asec, Swinyard et al.~\cite{Swin98}); the
pointing on SVS\,13 also includes HH\,11 and 10 and two sources VLA\,3
and SVS\,13B (see Sect.~\ref{intro}),  while that centered on HH\,7 also
includes HH\,8, 9 and 10.

\section{Results} 
\label{res}

Table~\ref{restab} presents the fluxes for the lines detected toward
each location. One sigma upper limits are given only for lines observed
with dedicated SWS02 ``line scans'' AOTs but not detected (see
Tab.~\ref{obstab}); a horizontal dash means that no observation is
available for that particular line.

\begin{table*}
\begin{center}
\caption{~~~~~~~~~~~~~~~~~~~~~~~~~~~~~~~~~~~~~~~~~~~~~~~~~~~~Observed Line Fluxes\label{restab}\tablenotemark{a}}
\vspace{0.25cm}
\begin{tabular}{lrrrrr}\tableline\tableline 
{Line} & {SVS\,13} & {HH\,7} & {HH\,10} & {Red Lobe 1} & {Red Lobe 2} \\ \tableline 
\multicolumn{6}{c}{SWS Lines} \\  \tableline  
(0-0)S(1)\tablenotemark{d} & 4.0(0.5) & 2.6(0.5) & 5.5(0.5) & 4.5(0.7) & 4.2(0.5) \\
(0-0)S(2)\tablenotemark{e} & 7.9(2.6) & 13.3(1.9) & 6.9(1.2) & 7.6(2.1) & 8.3(2.5) \\
(0-0)S(3)\tablenotemark{e} & 9.3(0.8) & 10.2(0.7) & 12.5(1.1) & 10.6(0.9)& 7.8(0.8) \\
(0-0)S(4)\tablenotemark{e} & 4.2(0.6) & 10.9(1.0) & 5.3(0.6) & 4.2(0.8) & 3.1(0.7) \\
(0-0)S(5)\tablenotemark{e} & 9.9(2.0) & 15.5(1.9) & 9.9(1.5) & 8.7(2.0) & 11.9(2.4)\\
(0-0)S(6)\tablenotemark{e} & $\leq$2 & $-$ & $\leq$3 & $-$ & $\leq$3\\
(0-0)S(7)\tablenotemark{e} & $\leq$4 & $-$ & $\leq$5 & $-$ & $\leq$5\\
\neii\tablenotemark{d} & $\leq$0.7 & $\leq$0.9& $\leq$0.7& $\leq$0.9  & 
$\leq$0.8 \\
\sii\tablenotemark{g} & 7.4(2.3) & 3.6(1.1) & 6.8(2.2) & 5.5(1.0) & $\leq$3 \\ \\ \tableline \\
\multicolumn{6}{c}{LWS Lines} \\  \tableline 
CO 20-19\tablenotemark{f}  & 5.4(1.7)\tablenotemark{b}& & $-$ & &  \\
CO 19-18\tablenotemark{f}  & 13.7(2.0)\tablenotemark{b} &  & $-$  &  & \\
CO 18-17\tablenotemark{f}  & 13.0(4.6)\tablenotemark{b} & & $-$ & & \\
CO 17-16\tablenotemark{f}  & 10.1(1.7)\tablenotemark{b} & 
8.4(3.0)\tablenotemark{c}& $-$ & \multicolumn{2}{c}{6.6(1.2)} \\
CO 16-15\tablenotemark{f}  & 10.1(1.7)\tablenotemark{b} & 
5.7(1.5)\tablenotemark{c} & $-$ & \multicolumn{2}{c}{3.9(0.6)} \\
CO 15-14\tablenotemark{f}  & 13.2(6.3)\tablenotemark{b} & 
8.5(2.8)\tablenotemark{c} & $-$ & \\
CO 14-13\tablenotemark{f}  & 10.1(2.7)\tablenotemark{b} & 
6.6(2.1)\tablenotemark{c} & $-$ & & \\
o-H$_2$O 3(0,3)-2(1,2)\tablenotemark{f} & 12.2(6.3)\tablenotemark{b} & 
6.5(2.8)\tablenotemark{c} & & & \\
o-H$_2$O 2(1,2)-1(0,1)\tablenotemark{f} & 6.8(3.8)\tablenotemark{b} & 
7.9(1.5)\tablenotemark{c} & & & \\
\oia\tablenotemark{f} & 180(2)\tablenotemark{b} & 131(3)\tablenotemark{c} & $-$ & \multicolumn{2}{c}{109(3)} \\
\oib\tablenotemark{f} & 8.0(4.0)\tablenotemark{b} & 4.6(1.0)\tablenotemark{c} & $-$ & \multicolumn{2}{c}{7.9(1.3)} \\
\cii\tablenotemark{f} & 21.3(1.4)\tablenotemark{b} & 12.5(1.1)\tablenotemark{c}
 & $-$ & \multicolumn{2}{c}{15.2(0.7)} \\ 
\tableline
\end{tabular} \\
\vspace{-1cm}
\tablenotetext{}{~~~~~~~~~~~~~~~$^a$ Units of 10$^{-20}$ \lflux and 1$\sigma$ 
uncertainties in parenthesis; upper limits are at the 1$\sigma$ level.}
\tablenotetext{}{~~~~~~~~~~~~~~~~~~A horizontal dash means that no observation 
is available for that particular line.}
\tablenotetext{}{~~~~~~~~~~~~~~~~$^b$ Includes HH\,10 and HH\,11.}
\tablenotetext{}{~~~~~~~~~~~~~~~~$^c$ Includes HH\,8 and; HH\,9 and HH\,10 are 
at the edge of the LWS beam.}
\tablenotetext{}{~~~~~~~~~~~~~~~~$^d$ Focal plane aperture is 14\asec\ x 
27\asec.}
\tablenotetext{}{~~~~~~~~~~~~~~~~$^e$ Focal plane aperture is 14\asec\ x 
20\asec.}
\tablenotetext{}{~~~~~~~~~~~~~~~~$^f$ Focal plane aperture radius is 40\asec.}
\tablenotetext{}{~~~~~~~~~~~~~~~~$^g$ Focal plane aperture is 20\asec\ x 
33\asec.}
\end{center}
\end{table*}

\begin{figure*}
\vspace{4cm}
\includegraphics{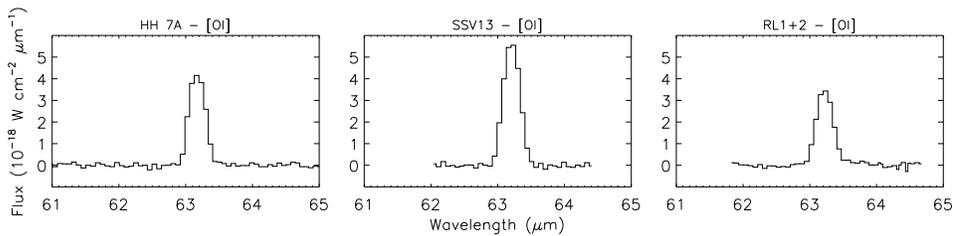}
\vspace{-1cm}
\caption{\label{oi63}{\bf a.} \oia\ lines detected in 
the LWS full  grating scans on the three observed positions (dashed 
circles in Fig.~\ref{map}).}
\end{figure*}

Detected lines are also plotted in Figs.~\ref{oi63}a, b, c and d; most of them 
are detected everywhere. Exceptions are \sii, which is not detected toward 
RL\,2, {\it ortho}-\ho\ which is not detected toward RL\,1+2, and \neii\ 
which is not detected anywhere. All lines, except \sii\ and the \hii\ lines, 
are stronger on the SVS\,13 position.

\addtocounter{figure}{-1}
\begin{figure*}
\vspace{8.5cm}
\includegraphics{fig_lws140-200.ps}
\caption[]{\label{lwslines}{\bf b.} Lines detected
between 130\um\ and 190\um\ in the LWS full  grating scans on the three
observed positions (dashed circles in Fig.~\ref{map}).}
\end{figure*}

\addtocounter{figure}{-1}
\begin{figure*}
\vspace{8cm}
\includegraphics{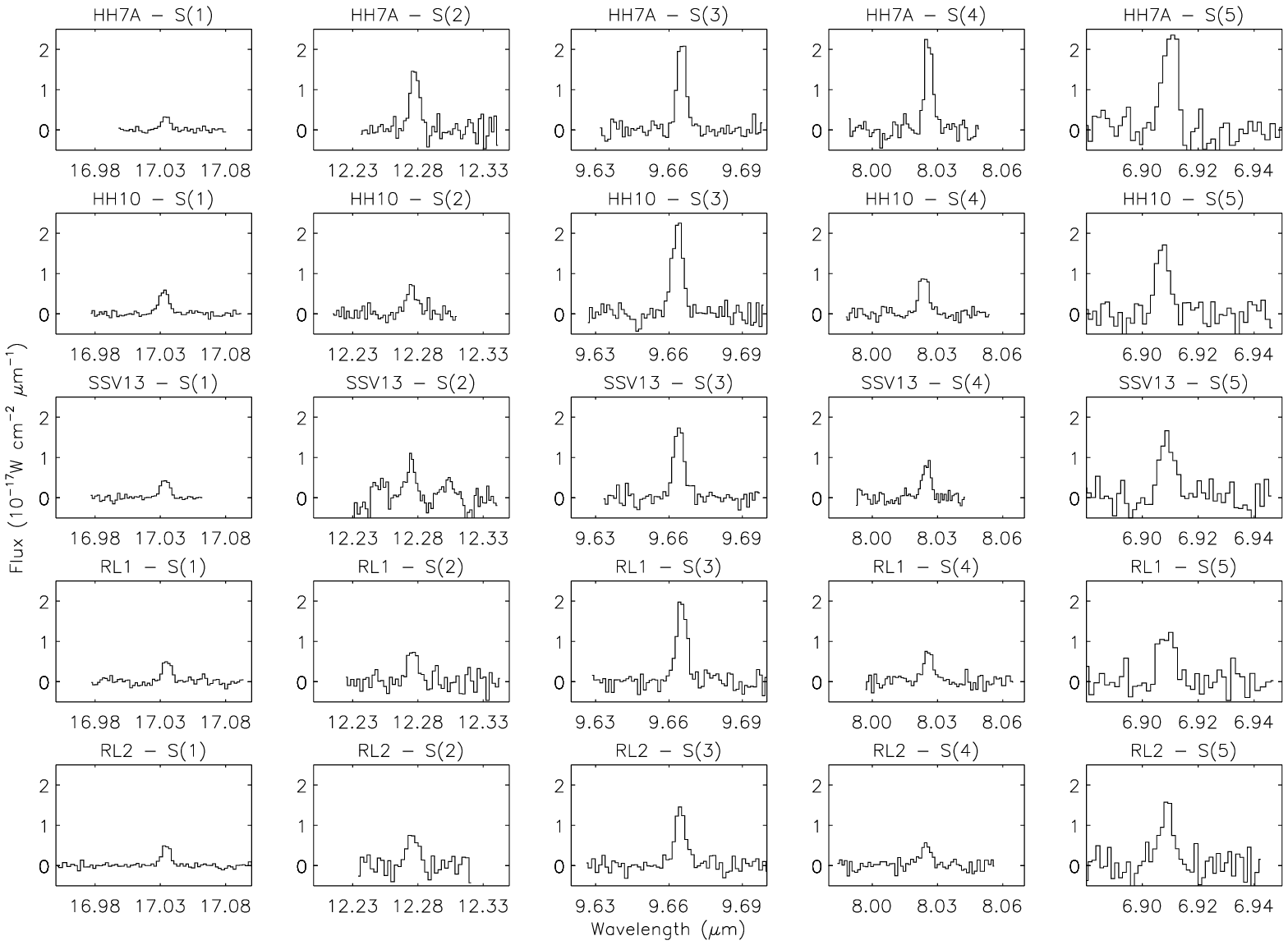}
\vspace{2cm}
\caption{\label{h2_lines}{\bf c.} \hii\ pure rotational
lines detected in the SWS line  grating scans on the five observed
positions (full rectangles in Fig.~\ref{map}).}
\end{figure*}

\addtocounter{figure}{-1}
\begin{figure*}
\vspace{3.cm}
\includegraphics{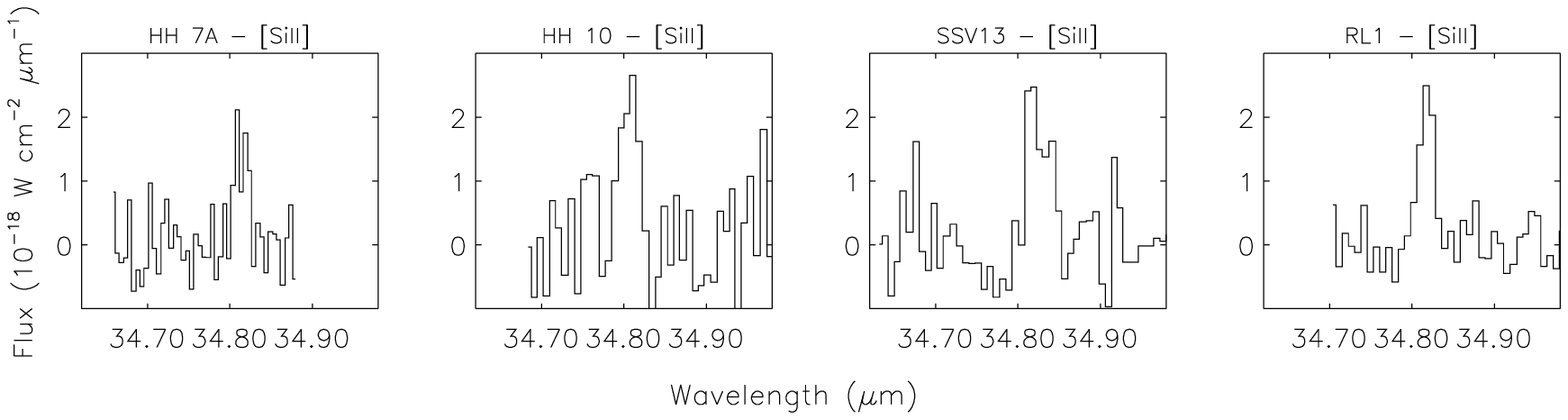}
\vspace{0.5cm}
\caption{\label{si2}{\bf d.} \sii\ lines detected in
the SWS line grating scans  on four out of five observed positions (full
rectangles in Fig.~\ref{map}).}
\end{figure*}

\label{dynamics}

\begin{figure*}
\vspace{4cm}
\includegraphics{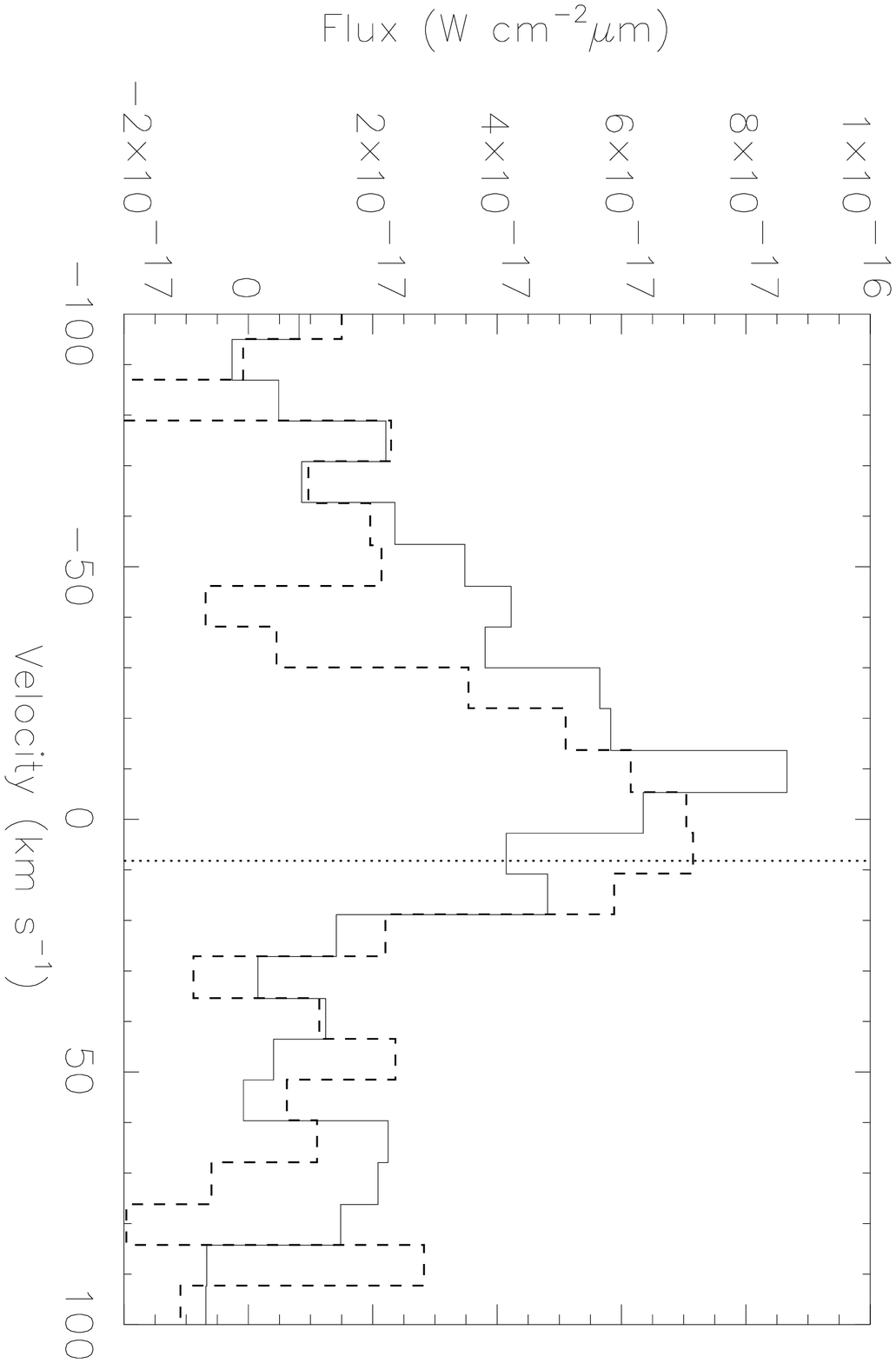}
\vspace{1cm}
\caption{\label{oi-fp}\oia\ line detected with the LWS
Fabry-Perot toward HH\,7  (full line) and RL\,1+RL\,2 (dashed line);
the dotted vertical line  indicates the cloud systemic velocity, 8.3
\kms.}
\end{figure*}

The FP spectra of the \oia\ line toward HH\,7 and RL\,1+2 are presented
in Fig.~\ref{oi-fp}. The lines are resolved ($\Delta v_{FP}\sim$30 \kms)
with deconvolved FWHM of $\sim$50 \kms and $\sim$25 \kms\ for the  two
positions; these widths  are consistent with the velocity field traced
by the [S~II]6717\AA\ line  (Stapelfeldt~\cite{S91}) toward HH\,7, and
suggest that \oia\ originates from the flow material.

\section{The Gas Physical Parameters}
\label{molecules}

\subsection{H$_2$}
\label{h2}

\begin{figure*}
\vspace{7cm}
\includegraphics{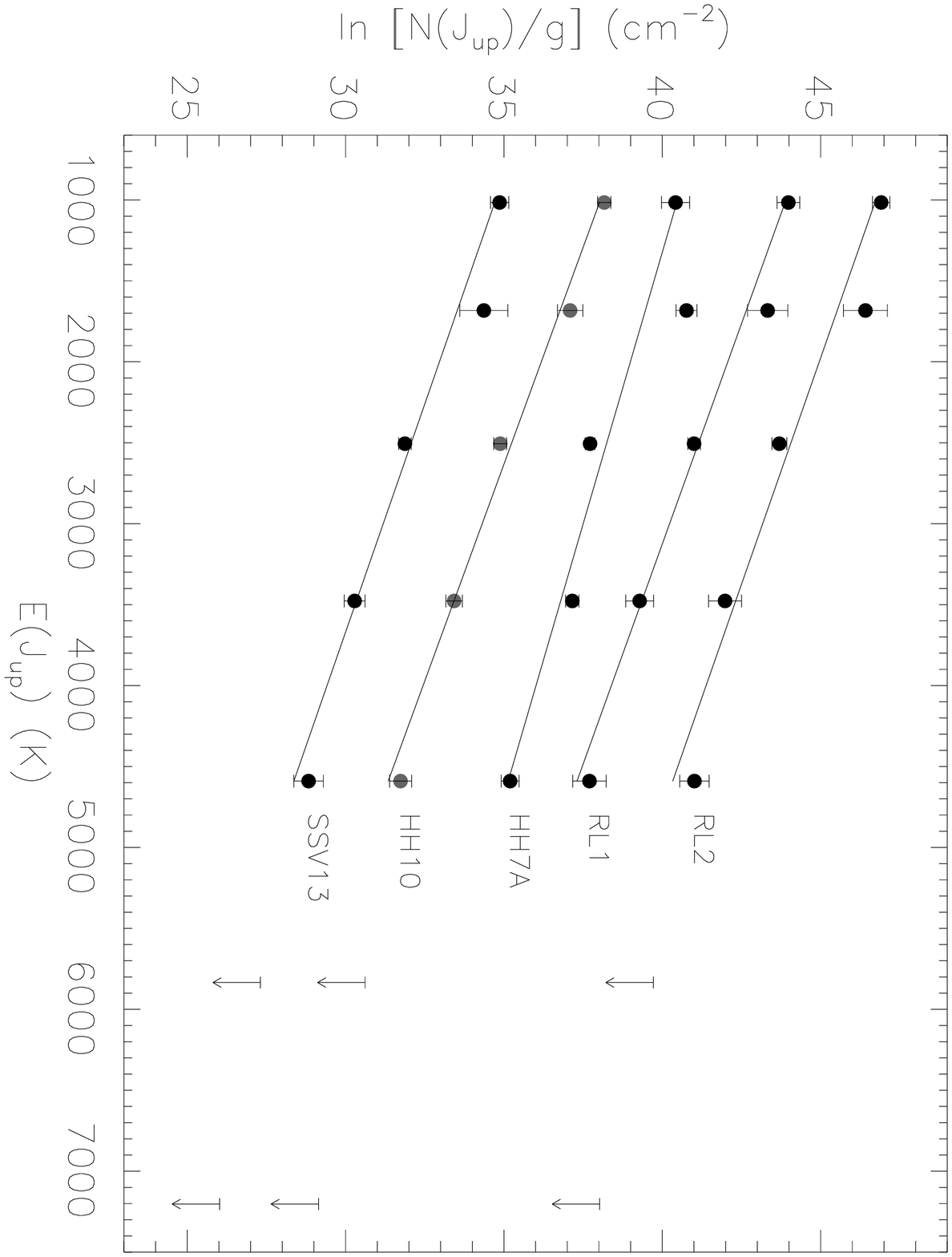}
\caption{\label{h2excite}H$_2$ excitation diagrams
for the five observed positions. In each panel, the abscissa represents
the energy of the level in K, while $N_j$/$g_j$ is in ordinata; offsets
of -10, -5, 5 and 10 dex have been applied to data of SVS\,13, HH\,10,
RL\,1 and RL\,2 while no offset has been  applied to the  HH\,7 data.}
\end{figure*}

The observed H$_2$ lines arise from quadrupole rotational transitions of
the ground vibrational state so that they are likely to be optically
thin. The radiative lifetimes range from 0.54 yrs for level $v$=0, J=7
where line (0-0)S(5) originates, to $\sim 1000$ yrs for level $v$=0, J=2
which is the upper state for (0-0)S(0) transition; thus these levels are
collisionally populated, and a simple LTE analysis is adequate to
interpret the data (e.g., Gredel~\cite{G94}). Einstein coefficients and
wavenumbers were taken from Black \& Dalgarno (\cite{BD76})  and
Dabrowski (\cite{D84}); ortho/para=3 has been assumed. We dereddened the
H$_2$ line fluxes using the visual extinction given by Gredel
(\cite{G96}) for the individual HH knots, and the Rieke \& Lebofsky 
(\cite{RL85}) extinction curve. Dereddened line fluxes were used to
produce the Boltzmann plots shown in Fig.~\ref{h2excite}; a single
temperature fit is in good agreement with the observed fluxes at all
positions. The solid angles of line emitting regions are arbitrarily set
to the equivalent SWS focal plane aperture,  i.e. 14\asec\ x 20\asec~or
6.6$\times10^{-9}$ sr (valid for all detected  \hii\ lines but S(1),
where the solid angle is $\sim$30\% higher), assuming a beam filling 
factor of 1.

The H$_2$ temperatures and the SWS beam averaged column densities,
together with 1$\sigma$ uncertainties are listed in
Table~\ref{physparam}. The \hii\ column densities should be considered
as lower limits because the beam filling factor can be less than 1. The
H$_2$ temperatures vary by less than 15\% along the flow and appear to
trace an H$_2$ component (which we call ``warm'') which is colder than
the one identified by Gredel (\cite{G96}) via higher excitation H$_2$
ro-vibrational lines (2100 K$\leq$ T $\leq$2750 K, which we will call
``hot''). The column densities for the warm H$_2$ (see
Table~\ref{physparam}, col. 3) are on average two orders of magnitude
higher than those of the hot H$_2$ (Gredel~\cite{G96}) and the
difference cannot be explained by the smaller (a factor 2) emitting
areas adopted by Gredel\footnote{We believe that Table 3 of Gredel
(\cite{G96}) incorrectly reports the \hii\ column densities; based on
the explanation in note $c$ of that Table, the total column densities
should be a factor ten higher than reported in its column 7. Furthermore
(R. Gredel, priv. comm.) the IRSPEC slit was not aligned exactly on the
peaks of  the individual knots, suggesting that quoted column densities
should more conservatively be considered as lower limits. 
\label{foot}}.  This suggests that the two H$_2$ components may be
actually distinct.

\begin{table*}
\begin{center}
\caption{~~~~~~~~~~~~~~~~~~~~~~~~~~~~~~~~~~~~~~~~~~~~~~~~~~~H$_2$ Physical Parameters\label{physparam}}
\vspace{0.25cm}
\begin{tabular}{lcc}\tableline\tableline 
{Object} & {T$_{\rm H_2}$} & {N(H$_2$)} \\
{~} & {(K)} & {(10$^{20}$ cm$^{-2}$)}\\  \tableline 
SVS\,13  & 560(20) & 0.7(0.1) \\
HH\,7    & 670(20) & 0.5(0.1) \\
HH\,10   & 540(10) & 1.0(0.1) \\
RL\,1   & 540(20) & 0.9(0.2) \\
RL\,2   & 560(20) & 0.8(0.1) \\  \tableline
\end{tabular}
\end{center}
\end{table*}

\subsection{CO}
\label{co}

CO rotational lines have been detected at all three positions observed
with LWS, although the intensity and number of detected lines are
highest toward SVS\,13. More CO lines than those reported in
Table~\ref{restab} are marginally visible in our spectra toward the two
flow positions, but it is impossible to reliably determine their flux 
because of the heavily fringed LWS spectra. Contrary to the \hii\ lines,
the CO rotational spectrum arises from dipole transitions and the
simplifying assumptions of optically thin lines cannot be adopted {\it a
priori}. We analysed the CO lines using the LVG model described in
Ceccarelli  et al. (\cite{Cetal98}). Under NLTE conditions the line
ratios depend both on the gas temperature and density, as well as on the
CO column density if the lines are optically thick. The observed CO
lines cannot constrain the three parameters simultaneously. The
distribution of CO line fluxes {\it vs} J toward SVS\,13 is essentially
flat, and we find a range of physical conditions which are consistent
with our observations. We find, as extreme cases, a `cold' solution with
T$\sim$350 K and $n\sim 10^6$ \cmthree, and a `warm' solution with
T$\sim$900 K and $n\sim 10^5$ \cmthree; the model also predicts
optically thin CO lines. Interestingly, this temperature range is
centered around the value independently derived from  H$_2$ lines, which
suggests that the CO and H$_2$ emission come from the same region. For
T=560 K (the H$_2$ temperature toward SVS\,13, see Table~\ref{physparam})
we obtain $n\sim 4\times 10^5$ \cmthree. The CO line ratios toward HH7
and RL\,1+2 imply temperatures and densities similar to those around
SVS\,13, although the lower absolute line fluxes would suggest column
densities or beam filling factor about a factor two lower. Using the
SWS beam solid angle we derive N(CO)$\sim 2\times 10^{16}$ \cmtwo\
toward the SVS\,13 LWS pointing, giving a {\it direct} measurement of
the CO abundance in this warm gas of [CO]/[H]$\sim 1.2\times 10^{-4}$.
Based on this number it seems that CO accounts for essentially all of
the gas-phase carbon in the interstellar medium  (Cardelli et
al.~\cite{CMJS96}). Recently Lefloch et al. (\cite{Letal98}) found
evidence of CO depletion toward the core of SVS\,13, likely due to
condensation onto grains in the dense environments close to SVS\,13. We
find no evidence of such a depletion in the high-J CO line emitting
region, suggesting that all of the CO locked into the grain mantles has
been returned to the gas-phase.

\subsection{H$_2$O}
\label{water}

As far as H$_2$O is concerned, since only the 3(0,3)-2(1,2) 174.6\um\
and 2(1,2)-1(0,1) 179.5\um\ lines of {\it ortho}-\ho\ have been detected
in SVS\,13 and HH\,7, it is clear that water plays only a secondary role
in gas cooling (see Table~\ref{cool-rates}) compared to some other cases
of outflow exciting sources such as IC\,1396N (Molinari et
al.~\cite{Metal98}) or L1448-mm (Nisini et al.~\cite{Netal99}). This
result goes against earlier suggestions based on previous $\lambda$=1.67
mm water line observations (Cernicharo et al.~\cite{CBG96}), that the
emission originates in dense (10$^6$ \cmthree) shocked material with a
water abundance comparable to that of CO.  It should be noted, however,
that Cernicharo et al. assumed a gas temperature of 50 K, while our
observations of \hii\ and CO lines clearly indicate that the temperature
of the molecular material is a factor of 10 higher. The maps presented
by Cernicharo et al. show that water emission is concentrated within 
$\sim 10$\arcsec~region centered on HH\,11. Adopting T = 560 K  and n =
4$\times 10^5$ \cmthree, compatible with the conditions for \hii\ and
CO, our model fit (Ceccarelli et al.~\cite{Cetal98}) to ISO and
millimeter water lines predicts  optically thick lines and a water
column density N(\ho)$\sim 10^{15}$ \cmtwo, about a factor 20 lower than
that estimated by Cernicharo et al. An abundance [H$_2$O]/[H$_2$]$\lsim6
\times 10^{-6}$ is also derived assuming that the lines from two
molecules come from the same gas component; we will prove this
assumption in Sect.~\ref{shocks}.

\section{The Photo-Dissociation Region}
\label{pdr}

In spite of the shock-excited nature of the HH objects, it is likely
that a non-negligible contribution to the observed line emission
actually come from an extended Photo-Dissociation Region (PDR, Tielens
\& Hollenbach~\cite{TH85}) component associated with the NGC\,1333
cloud. This is suggested by low dispersion ISO-LWS observations (Caux et
al.~\cite{Cetal00}) at various positions in  the NGC\,1333 cloud, which
show a widespread \oia\ and \cii\ emission; this extended component
seems to account for $\sim$20\% of the \oia, and all of the \cii\
emission we see from our LWS pointings. Our FP data
(Sect.~\ref{dynamics}) confirm that only a small fraction of the \oia\
line may come from a quiescent PDR  component. 

Using the {\it PDR Toolbox}\footnote{The {\it ``PDR Toolbox''} is
available at http://dustem.astro.umd.edu and contains downloadable FIR
lines diagnostic information about PDRs. The tool has been created by L.
Mundy, M. Wolfire, S. Lord and M. Pound, and it is based on the new PDR
models of Kaufman et al. (\cite{KWHL99}).\label{toolbox}}, the observed
\cii\ emission requires a relatively faint FUV irradiation level of
G$_0\lsim$10 in units of local Galactic FUV flux (Habing~\cite{H68}).
Although SVS\,13 might be able to provide the required FUV field, the
widespread \cii\ emission seen by Caux et al. (\cite{Cetal00}) clearly
suggest an external irradiation source. A natural  candidate is BD
+30$^{\circ}$\,549, the B6 star responsible for the illumination of the
NGC\,1333 reflection nebulosity (Harvey et al.~\cite{Hetal84}); located
$\sim$ 0.8 pc N-NE of the HH\,7-11 area, it can certainly provide the
needed G$_0\lsim 10$ FUV field. In this regime, the PDR surface
temperature does not exceed $\sim$ 100 K (Kaufman et al.~\cite{KWHL99},
Timmermann et al.~\cite{Tetal96}, Kemper et al.~\cite{Ketal99}, Liseau
et al.~\cite{Letal99}) which excludes a PDR origin for the \sii\ line
(Hollenbach, Takahashi \& Tielens~\cite{HTT91}), the \hii\ and the CO
lines. Indeed, the temperature of the molecular material is at least 5
times higher, and the molecular emission does not appear to be extended
as one would expect for a PDR origin; ISOCAM-CVF near-IR imaging
spectroscopy of the HH\,7-11 region extracted from the public ISO data
archive\footnote{The ISO archive is available at
http://pma.iso.vilspa.esa.es\label{isoarchive}} shows that the emission
from the same \hii\ rotational lines observed with SWS is concentrated
along the flow and peaks in correspondence of visible HH objects
(Noriega-Crespo et al.~\cite{Nor00}), obviously favouring a shock origin
for these lines.

\section{Shocks along the Flow}
\label{shocks}

\begin{table*}
\begin{center}
\caption{~~~~~~~~~~~~~~~~~~~~~~~~~~~~~~~~~~~~~~~~~~~~~~~~~~~~~~~~~~Cooling rates\tablenotemark{a}\label{cool-rates}}
\vspace{0.25cm}
\begin{tabular}{lccccc}\tableline\tableline 
{Species} & {HH\,7} & {HH\,10} & {SVS\,13} & 
{RL\,1} & {RL\,2} \\ \tableline 
\oia   & 4.8 &  & 6.6 & \multicolumn{2}{c}{4.0} \\
CO     & 2.0 &  & 3.0 & \multicolumn{2}{c}{1.4} \\
H$_2$O & 0.7 &  & 0.7 & \multicolumn{2}{c}{$-$} \\
\hii & 2.3 & 1.9 & 1.6 & 1.7 & 1.6 \\
\sii & 0.13 & 0.25 & 0.27 & 0.20 & $-$\\  \tableline
\end{tabular} \\
\vspace{-1cm}
\tablenotetext{}{~~~~~~~~~~~~~~~~~~~~~~~~~~~~~~~~~~~~~~~~$^a$ Units of 10$^{-2}$ \lsun.}
\end{center}
\end{table*}

It is known that the nature of the shock excitation is dramatically
influenced by the presence of a magnetic field component  perpendicular
to the shock velocity which prepares the up-stream medium and smoothes
out the effect of the front passage. The differences in the physical
conditions of shocked gas are such that two distinct classes of shocks,
C({\it ontinuous}) and J({\it ump}) have been idealised
(Draine~\cite{D80}). In a J-shock the temperature reached by the shocked
material depends on the square of the shock velocity and can be as high
as 10$^5$ K, resulting in complete molecular dissociation. In C-shocks,
the temperature rarely exceeds a few thousand degrees and  molecular
material can survive. These two very different physical scenarios
produce distinctive signatures in terms of cooling ratios between
different species, and of line ratios within the same species. We will
show that our data of the HH\,7-11 region depict a complex situation
where the two types of shock coexist; to help the discussion below, we
report in Table~\ref{cool-rates} the total cooling rates in the various
species as, when applicable,  derived from the models used to estimate
their physical parameters. 

\subsection{J-shocks}
\label{jshocks}

First of all we note that the non-detection of the \neii\ line down to a
level of 10$^{-20}$ \lflux\ confirms the low-excitation nature of the
HH\,7-11 chain and suggests (Hollenbach \& McKee~\cite{HM89}, hereafter
HM89) a shock velocity v$_s\lsim 40-50$ \kms, depending of the pre-shock
density, in excellent agreement with our high-resolution FP \oia\
observations (Sect.~\ref{dynamics}) and with estimates from optical
spectroscopy (B\"ohm, Brugel \& Olmsted~\cite{BBO83}; Solf \&
B\"ohm~\cite{SB87}). This upper limit on the shock velocity excludes
strong shocks, yet the detection of \sii\ requires the presence of a
dissociative shock component, since negligible ionization is expected
from a C-shock (HM89). Such a component would also explain the observed
\oia\ cooling, since the latter is expected to be the main coolant in
J-shocks\footnote{Copious \oia\ can also be produced in non-dissociative
shocks. In this case however, the presence of \hii\ would allow the
incorporation of O into water, via the chain of endothermic reactions O
+ \hii\ $\rightarrow$ OH + H and OH + \hii\ $\rightarrow$ \ho\ + H. This
chain has an activation energy of T$\sim$ 220 K, lower than the \hii
temperature estimated from the rotational lines (between 540 and 670 K,
see Table~\ref{physparam}); we conclude that the observed \oia\ emission
cannot originate in C-shocks.}. In order to compare the \oia\ with the
\sii\ line for HH\,7 and SVS\,13, we need to determine the fraction of
\oia\ emission due to other HH objects falling within the LWS beam.

The LWS beam centered on HH\,7 also contains HH\,8, 9 and 10; since the
excitation conditions for the different HH objects along the flow do not
show dramatic variations (Hartigan, Curiel \& Raymond~\cite{HCR89}), we
choose to use the detected \sii\ lines toward  HH\,7 and HH\,10 as
weights to estimate the \oia\ cooling intrinsic to HH\,7. We
conservatively assign to HH\,8 and HH\,9 the same \sii\ flux measured
toward HH\,7, and we also consider that the weighting of the LWS beam
profile decreases the  contribution of HH\,8-9 and HH\,10 to the \oia\
measured on the HH\,7 pointing by 10\% and 50\%, respectively. Similar
estimates can be done for the SVS\,13 pointing, where again HH\,10
contributes at a 50\% level. The HH\,11's contribution does not need to
be disentangled since it also contributes to the \sii\ line of SVS\,13.
No correction is required for the RL\,1+2 pointing. Taking the above
into account, we obtain \oia/\sii\ line ratios between 15 and 20,
suggesting a pre-shock density $n_0 \sim 10^4$ \cmthree\ (HM89). The
models can also reproduce the absolute fluxes, provided that the
emission solid angles are $\sim$ 5\asec\ in diameter with filling factor
of 1; more intense O{\sc i} and  Si{\sc ii} lines  on  the central
position are likely due to slightly  larger solid angles and/or filling
factors.

As concerns molecular cooling the main contribution to \hii\ emission in
J-shocks is predicted (HM89) to come from material excited by FUV  or 
\hii-formation pumping. Indeed, Fernandes \& Brand (\cite{FB95}) propose
a fluorescent origin for the near-IR \hii\ lines toward HH\,7. For the
(0-0)S lines however, the predicted line ratios are not  reproduced by
the observations. In particular the S(7) line, which is predicted to be
always brighter than the S(5) line irrespectively of pre-shock density
and shock velocity, is not detected at all in our SWS
spectra\footnote{The above mentioned ISOCAM-CVF observations (at lower
spectral resolution than the SWS) actually detected the S(7) but with a
flux about 4 times lower than the S(5) line flux, compatibly with our
SWS upper limits on the S(7) line (Noriega-Crespo et
al.~\cite{Nor00}).}.

\subsection{C-shocks}
\label{cshocks}

\begin{figure*}
\vspace{7cm}
\includegraphics{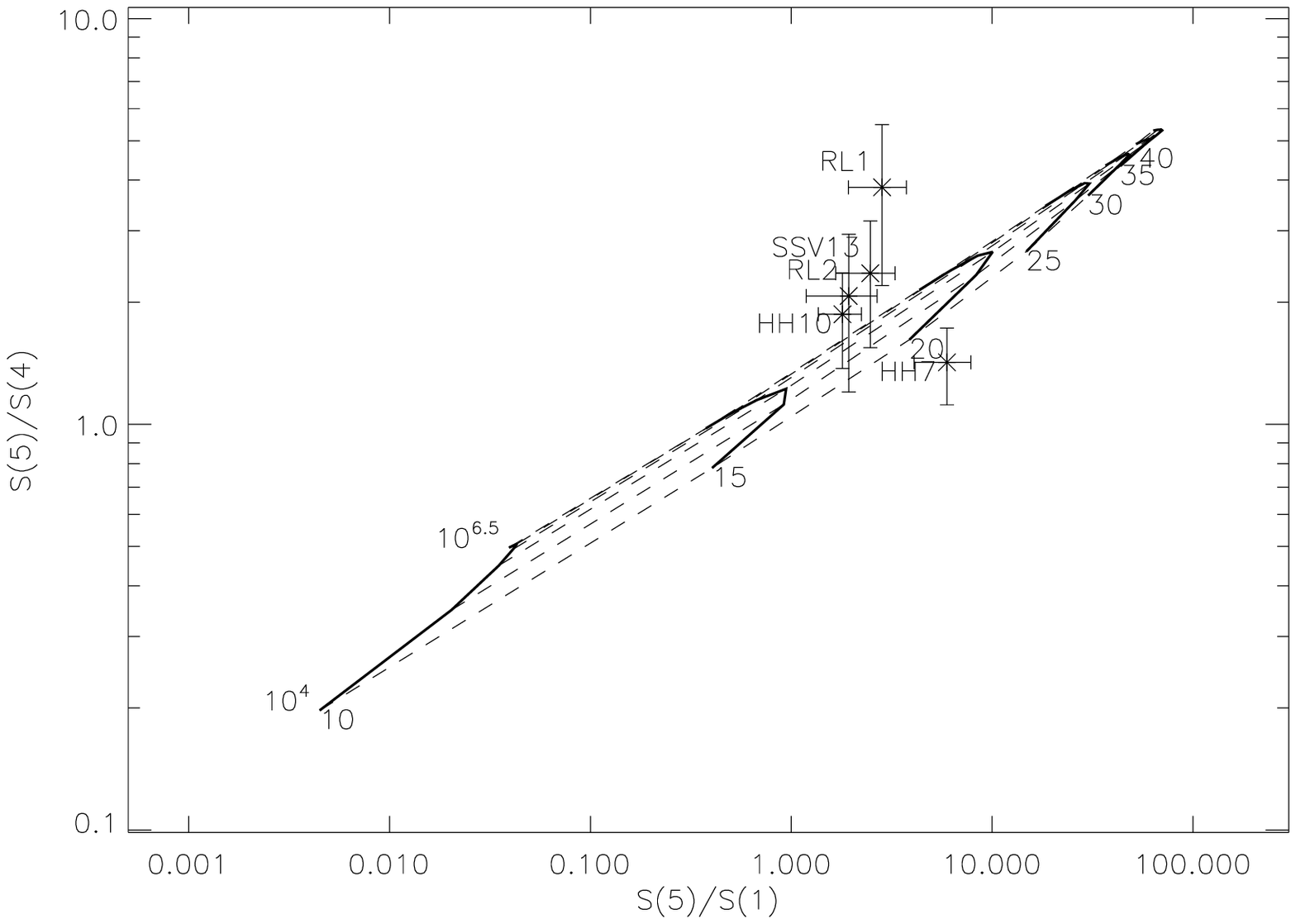}
\caption{\label{h2_diagc} C-shock diagnostic ratios for
the \hii\ pure rotational lines. The solid and dashed lines represent
iso-density and iso-velocity levels respectively. The levels are labeled
in \cmthree\ (powers of ten) and in \kms. The crosses represent the
\hii\ line ratios for the five positions observed with the SWS.}
\end{figure*}

When interpreted in terms of non-dissociative shock, the pure rotational
\hii\ lines provides a sensitive probe for the shock velocity, given the
four orders of magnitudes of dynamical range spanned by their line
ratios. This is shown in Fig.~\ref{h2_diagc}, which presents the
observed \hii\ line ratios superimposed on a grid of C-shock models from
Kaufman \&  Neufeld (\cite{KN96}). 

\begin{figure*}
\vspace{7cm}
\includegraphics{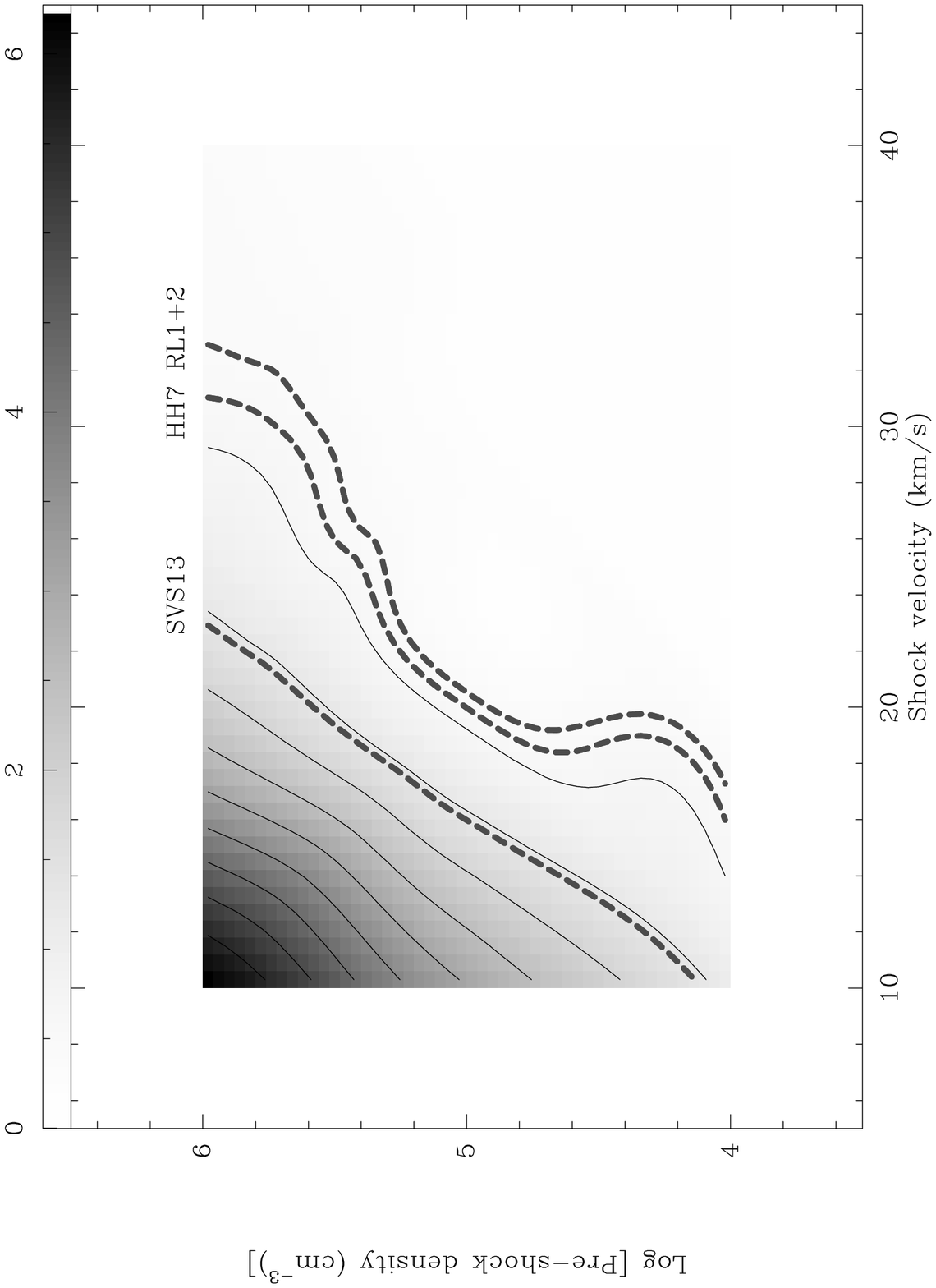}
\caption{CO/\hii\ cooling ratio as predicted for
C-shock models (Kaufman \& Neufeld 1996) as a function of pre-shock
density and shock velocity. The dashed lines indicate the values
observed in the different LWS pointings.\label{ccoh2ratio} }
\end{figure*}

\begin{figure*}
\vspace{10cm}
\includegraphics{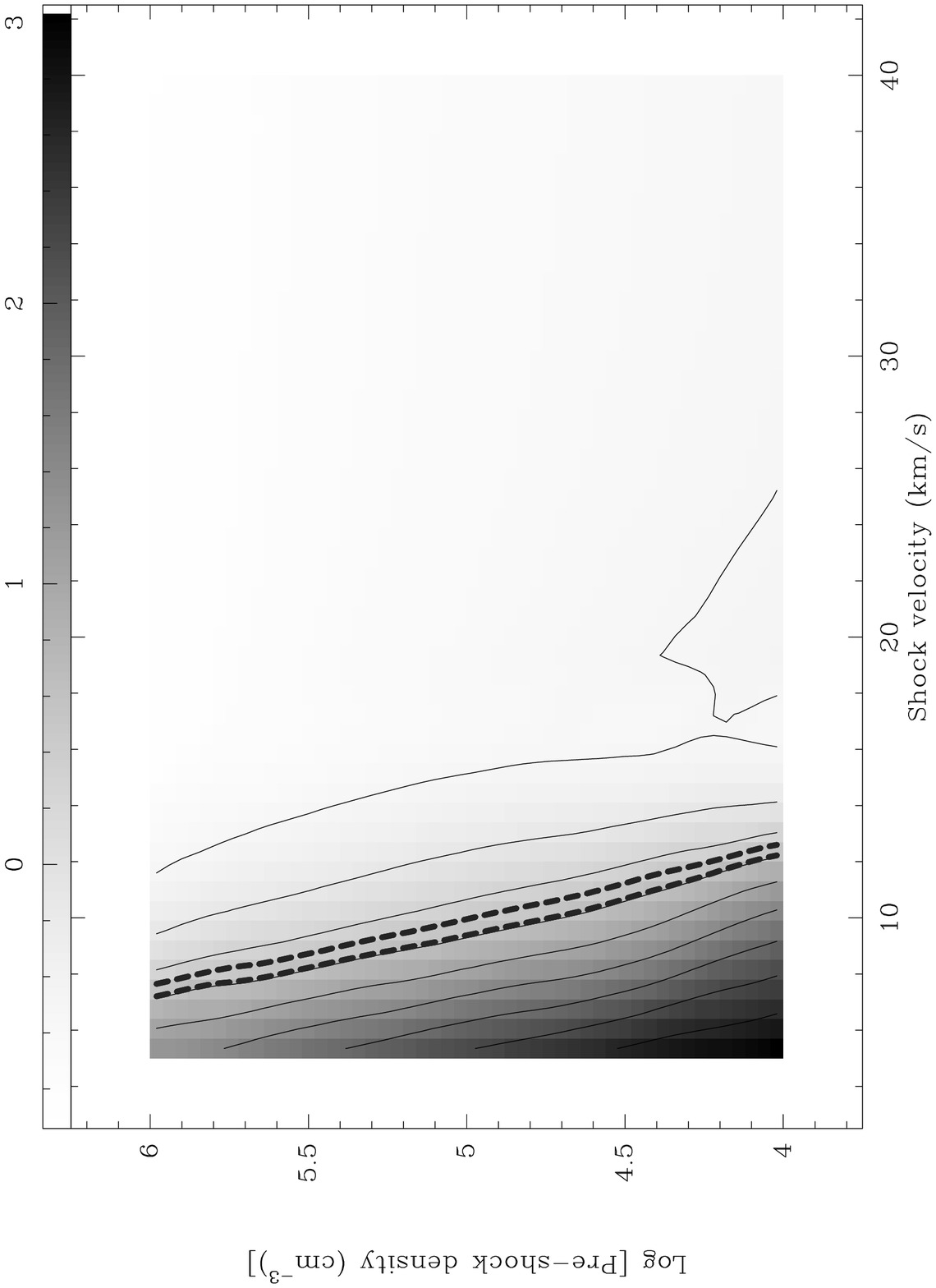}
\caption{Decimal logarithm of the CO/\ho\
cooling ratio as predicted for C-shock models (Kaufman \& Neufeld 1996)
as a function of pre-shock density and shock velocity. The dashed lines
indicate the values observed toward HH\,7 and SVS\,13 (water emission
was not detected toward RL\,1+2).\label{ccoh2oratio}}
\end{figure*}

Our observations seem to place quite stringent boundaries to the shock
velocity, $15\lsim v_s \lsim 20$ \kms. The corresponding gas
temperature, $\sim 600$ K (Kaufman \& Neufeld~\cite{KN96}), is in good
agreement with the values derived using simple LTE analysis
(Sect.~\ref{molecules}), as expected since the molecular gas is
collisionally heated by the C-shock front. Observed  absolute fluxes can
also be reproduced by the model as long as the emission solid angle
does  not exceed few arcsecs in diameter. We note that such a low
velocity  shock would also produce very faint S(6) and S(7) lines.  Once
the shock velocity is determined, we can use the CO/\hii\ cooling ratio
to estimate the pre-shock density. Fig.~\ref{ccoh2ratio} presents such a
diagnostic diagram.

The CO line fluxes in the three LWS pointings should be corrected for 
contamination by the other HH objects. In similar way as for the
\oia/\sii\ ratio, we use the \hii\ coolings in different positions as
weights to split the individual contributions to the CO cooling. 
The corrected cooling ratios are reported as dashed lines in
Fig.~\ref{ccoh2ratio}. We see that a shock velocity $15\lsim v_s \lsim
20$ \kms\ would correspond to a pre-shock densities of $n_0\sim 10^4$
\cmthree\ for HH\,7 and RL\,1+2, and $n_0\sim 10^5$ \cmthree\ for
SVS\,13. 

Water is an important diagnostic for C-shock models. \ho\ cooling is a
fast function of the shock velocity since free atomic oxygen in the
post-shock gas is expected to be incorporated into water as soon as the
temperature rises above $\sim200$ K, roughly corresponding to $v_s\sim
10$ \kms\ (Draine, Roberge \& Dalgarno~\cite{DRD83}, Kaufman \&
Neufeld~\cite{KN96}). In our case then, where the temperature is
$\sim$600 K,  cooling via H$_2$O rotational transitions is expected to
be dominant with respect to that of CO. Instead, the diagnostic diagram
in Fig.~\ref{ccoh2oratio} shows that the observed CO/\ho\ cooling ratio
is consistent with shock velocities $v_s\lsim 12$ \kms, which cannot
justify the observed \hii\ temperatures. This suggests that most of the
expected gas phase water is  missing. We will propose an explanation for
this result in Sect.~\ref{hh7}

\section{The Shocks and the Herbig-Haro Objects}
\label{shockshh}

We have shown that HH\,7-11 is a complex region where  different line
emission mechanisms are simultaneously at work.   Notwithstanding the
poor spatial resolution of our data, the variety of  different spectral
signatures detected allow us to draw the following physical scenario of
the region.

\subsection{HH\,7}
\label{hh7}

Starting from HH\,7, the C-shock conditions diagnosed by the molecular
emission clearly require the presence of a magnetic field
\bo$\perp$\vsc\ (Draine~\cite{D80}),  where \vsc\ is the shock velocity
of the C component. Assuming the standard  scaling law
(Draine~\cite{D80}) for the value of \mbo\,

\begin{equation}
B_0 \sim b~ n_0^{0.5}~~\mu{\rm G}
\label{magnetic}
\end{equation}

where $b\sim1$ in the interstellar medium, the estimated $n_0\sim~
10^4$\cmthree\ pre-shock density for the C-shock component on HH\,7
gives \mbo$\sim100$\mug. Variations of a factor 3 in each direction are
nevertheless possible, since $b$ can vary between 0.3 and 3 in molecular
clouds (HM89). For similar \mbo\ and $n_0$, Draine, Roberge \& Dalgarno
(\cite{DRD83}) have shown that the C$\rightarrow$J
transition occurs at \mvs$\sim$50 \kms, similar to the upper limit set
on \mvsj\ (the J-shock velocity) by our observations. This means that a
significant \bo\ component transverse to \vsj\ would smooth the shock
front to a C type or, equivalently, that our observed J-shock component
can only exist as long as \vsj$\parallel$\bo.  When the bowshock-like
morphology of HH\,7 is also considered, as revealed by HST NICMOS
2.12\um~images (A. Cotera, priv. comm.), then  a very simple scenario
emerges where HH\,7 is immersed in a \bo\ roughly parallel to the flow
axis. At the tip of the bow \vs$\sim \parallel$\bo\ and J-shock
conditions are present; \sii\ and most of \oia\ flux arise from this
region, and favourable conditions also exist to originate the ``hot'',
T$\sim2100$ K, \hii\ emission detected by Gredel (\cite{G96}).  Along
the sides of the bow, \vs\ becomes nearly perpendicular to \bo, 
creating favourable conditions for C-type shocks. This picture also
provides a plausible explanation to the problem of the missing water:
the water is actually produced in the C-shocks, but rapidly  condenses
onto dust grains and disappears from the gas-phase. This possibility is
suggested by the recent discovery with ISO-LWS (Molinari et
al.~\cite{Metal99}) of crystalline water ice toward HH\,7. The deduced
water abundance (in solid state form) is comparable to the interstellar
oxygen abundance, which is what the models would predict for gas-phase
water in C-shocks; furthermore, the fact that the ice is in crystalline
form requires grain temperatures of the order of 100 K, only attainable
in dissociative shocks (HM89; Draine, Roberge \& Dalgarno~\cite{DRD83}).
After being heated, the grains would be efficiently transported in the
post C-shock regions along the \bb\ lines, directed parallel to the flow.
For the gas parameters we derived, the grains are efficiently coupled to
the magnetic  field and are not significantly decelerated by collisions
with the  neutrals (Draine, Roberge \& Dalgarno~\cite{DRD83}). 

\subsection{HH\,10}
\label{hh10}

From the viewpoint of the line emission properties, HH\,10 and HH\,7
appear very similar objects. Although no LWS data were specifically
collected toward HH\,10, the \hii\ and the \sii\ lines still argue in
favour of a dual J+C shock nature. From the morphological point of view,
HH\,10 appears as an irregular blob in  the 2.12\um\ image of 
Fig.~\ref{map}. Higher spatial resolution images in  \ha\ and \suii\
from HST (unpublished archival data) resolved HH\,10 into a  double
filamentary structure whose N-S orientation does not appear to be 
related to the flow direction. However, the \ha/\suii~emission ratio is
higher in the NW part of this HH object, facing toward SVS\,13. A higher
ratio implies higher excitation (stronger shocks) conditions, which are
likely to be traced by the \sii\ line and by the `hot', T$\sim2200$ K, 
\hii\ component (Gredel \cite{G96}). If we assume that  \bo\ 
maintains its direction downstream along  the flow, as suggested for
HH\,7 (see above), then we speculate that the morphology of HH\,10
could  correspond to irregularities or corrugations in the shocked walls
of the cavity which is excavated by the flow. Such irregularities might
originate from instabilities at the flow-cavity interface, as proposed
by Liseau, Sandell \& Knee (\cite{LSK88}), although there are no
detailed numerical simulations of the process.

\subsection{HH\,11 and SVS\,13}
\label{hh11}

The interpretation for HH\,11 and SVS\,13 is more complicated because
both sources are contained in the fields of view of the SWS and LWS
instruments. Pre-shock densities $n_0\sim10^5$ \cmthree\ are here found
for the C component; this is a factor ten higher than in the other 
positions of the flow, which is not surprising given the  close 
proximity of the origin of the flow. In these conditions the magnitude 
of the magnetic field (Eq.~\ref{magnetic}) could range  from
$\sim300$\mug\ to $\sim$1mG. Such high \mbo\ values have been  claimed
by Hartigan, Curiel \& Raymond (\cite{HCR89}) to justify the  relative
faintness of HH\,11 in 2.12\um\ \hii\ images (see also  Fig.~\ref{map}).
The possibility that FIR line emission toward SVS\,13 arises in a
collapsing envelope around  is not relevant here because the predicted
\oia\ line  flux (Ceccarelli,  Hollenbach \& Tielens~\cite{CHT96}) is
about 30 times lower than  actually observed, while CO and \ho\ lines
are below the detection  limit for the present observations. Finally,
there is the possibility that a fraction of the line fluxes measured
with the LWS originates from the recently discovered embedded outflow
source SVS\,13B (Bachiller et al.~\cite{Betal98}).

\subsection{The Red-shifted Lobe}
\label{redlobe}

\oia, \oib\ and \sii, together with a complement of \hii\ and CO lines,
have been detected toward the receding lobe, and define shock conditions
which are not dramatically different from those present on  the blue
lobe. It is well known that  no optical emission is detected  toward the
red lobe, and also published images in the \hii\ 2.12\um\  line
(Fig.~\ref{map}, Garden et al.~\cite{Getal90}, Hodapp \&
Ladd~\cite{HL95}) clearly show fainter emission there. Higher values of
dust extinction with respect to the blue lobe have been invoked as an
explanation for this asymmetry. Our observations, which trace similar
shock conditions for the two lobes, tend to support this possibility.

\section{The Shocks and the Molecular Outflow}
\label{shocksoutflow}

If J-shocks between stellar winds and ambient material are responsible
for the acceleration of the molecular outflow, a correlation
(Hollenbach~\cite{H85}) is expected between the outflow mass loss rate
and the flux of the \oia\ line, which is the dominant coolant in such
shocks. The predicted mass loss rate, based on the observed \oia\
cooling from HH\,7, is $4.8\times 10^{-6}$ \msunyr\ (see also Cohen et
al.~\cite{Cetal88}, Ceccarelli et al.~\cite{Cetal97}), in good agreement
with the mass loss rate  estimated by Lizano et al. (\cite{Letal88}, see
also Rodriguez et  al.~\cite{Retal90}) for the fast H{\sc i} wind
believed to be responsible  for the acceleration of the slow CO outflow
(Snell \& Edwards~\cite{SE81}) . This fast neutral wind was  also
confirmed with CO observations by Bachiller \& Cernicharo (\cite{BC90}).

Assuming momentum balance at the interface between the the wind and the 
ambient medium, Davis \& Eisl\"offel (\cite{DE95}, \cite{DE96}) derived
a simple relationship between the mechanical power of the wind $L_w$
and  the power radiated by the shock $L_{rad}$:

\begin{equation}
{{L_{rad}}\over {L_w}} = {{v_s}\over {v_w}}\left[1-{{v_s}\over {v_w}}\right]^2
\label{balance}
\end{equation}

Since the working surfaces where the winds impact the medium are traced
by the J-shocks, we make the assumption $L_{rad}\sim L_{O{\sc 
i}}+L_{Si{\sc ii}}+L_{H_2-rovib}$. The total cooling due to the near-IR
\hii\ vibrational lines measured by Gredel (\cite{G96}) along the flow
is $\sim 5\times 10^{-3}$ \lsun; we assume a slit width of few arcsecs,
so we will conservatively multiply the observed value by 10 to allow for
the extension of the HH objects. We estimate the mechanical power of the
H{\sc i} wind according to:

\begin{equation}
L_w={{1}\over {2}} {{M_w v_w ^2}\over {\tau _{dyn}}}
\label{lumkin}
\end{equation}

Using the H{\sc i} parameters from Lizano et al. (\cite{Letal88}), we
obtain $L_w\sim 1.9$ \lsun. Eq. (\ref{balance}) then provides 
$v_s/v_w\sim 0.6$, or \mvsj$\sim 36$ \kms\ for an average H{\sc i} wind
velocity of $v_w\sim$60 \kms\ (Lizano et al.~\cite{Letal88}), in
excellent agreement with our results.

\section{Summary}
\label{summary}

The HH\,7-11 flow, together with its red-shifted counterpart and SVS\,13
(the candidate exciting source) have been studied via atomic, ionic and 
molecular spectroscopy. A complex scenario emerges, where:

\begin{enumerate}

  \item we have detected atomic (\oia, \oib), ionic (\cii, \sii) and 
molecular (\hii, CO and \ho) lines along the flow (both lobes) and
toward SVS\,13.

  \item the low-excitation shock nature of the HH nebulosities along the
flow is confirmed. Spectral signatures of C and J shocks are
ubiquitously found along the HH\,7-11 flow and its red-shifted
counterpart. Our estimates for the shock velocities are
\mvsj$\lsim40-50$ \kms\ and 15$\lsim$\mvsc$\lsim$20 \kms. The pre-shock
density is $\sim10^4$ \cmthree\ toward the blue and the red lobe; for
the C  component only, we find $n_0\sim 10^5$ \cmthree\ at the location
of  SVS\,13.

  \item there is indirect evidence for an ordered \bb\ field oriented
parallel to the direction of the flow. The magnitude of the magnetic  
field is \mbo$\sim100$\mug\ on the lobes, increasing to $\sim300$\mug\  
at the position of the flow origin; these figures, however, can vary
of a factor 3 in each direction.

  \item the gas-phase in the post C-shock region is deficient in \ho. 
We presented evidence that this may be due to freezing onto warm grains 
processed through the J-shock front and traveling downstream along the 
magnetic field lines.

  \item the asymmetry in optical and NIR properties among the two  lobes
of the outflow is probably not caused by different pre-shock densities
or shock velocities, supporting the hypothesis of higher extinction 
values toward the red lobe.
  
  \item the total J-shock cooling is compatible with the molecular
outflow being accelerated by the fast neutral wind detected in H{\sc i}
and CO.
  
  \item the whole flow area appears to be associated with a faint PDR
illuminated by BD +30$^{\circ}$\,549, the source responsible for the
illumination of the whole NGC\,1333 nebula.

\end{enumerate}


We thank L. Testi and M. Cecere for their  assistance with the
observations and data reduction of the \hii\  2.12\um\ image presented
in Fig.~\ref{map}. The staff of the Mt.  Palomar 60\asec\ telescope is
also acknowledged. We also thank an anonymous referee whose comments
improved the paper, and L.F. Rodriguez for his  comments on an early
version of this  manuscript. The ISO Spectral Analysis Package  (ISAP)
is a joint development by the LWS and SWS Instrument Teams and  Data
Centers. Contributing institutes are CESR, IAS, IPAC, MPE, RAL  and
SRON.

%
%
%
%
%
%

\end{document}